\documentclass{appolb}

\usepackage{graphicx}
\usepackage{subfigure}
\usepackage[cp1250]{inputenc}
\usepackage[colorlinks=true,linktocpage=true,linkcolor=blue,citecolor=blue]{hyperref}
\usepackage[usenames,dvipsnames]{color}

\def\ba{\begin{eqnarray}}
\def\ea{\end{eqnarray}}
\def\lb{\label}
\def\nn{\nonumber \\}

\def\d{\delta}

\def\rr{\rightarrow}

\def\e{\eta}

\def\t{\tau_f}

\def\l{\lambda}

\def\p{\perp}

\def\sp{\;\;\;\;}
\def\th{\theta}

\begin{document}

\title{Bose-Einstein correlations and thermal cluster formation in high-energy  collisions 
\thanks{email addresses: bialas@th.if.uj.edu.pl, wojciech.florkowski@ifj.edu.pl,  and \\ zalewski@th.if.uj.edu.pl}
}
\author{Andrzej Bialas
\address{M. Smoluchowski Institute of Physics, Jagellonian University, PL-30-059~Krakow, Poland}
\and
Wojciech Florkowski
\address{The H. Niewodnicza\'nski Institute of Nuclear Physics, Polish Academy of Sciences, PL-31342 Krak\'ow, Poland, and \\
Institute of Physics, Jan Kochanowski University, PL-25406~Kielce, Poland}
\and
Kacper Zalewski
\address{M. Smoluchowski Institute of Physics, Jagellonian University, PL-30-059~Krakow, Poland, and \\
The H. Niewodnicza\'nski Institute of Nuclear Physics, Polish Academy of Sciences, PL-31342 Krak\'ow, Poland
}
}
\maketitle

\begin{abstract}
The blast wave model is generalized to include the production of thermal clusters, as suggested by the success of the statistical model of particle production at high energies. The formulae for the HBT correlation functions and the corresponding HBT radii are derived.
\end{abstract}

\section{Introduction}  

The soft hadronic data collected in high energy collisions are frequently analyzed in the framework of {\it thermal} or {\it statistical models} (see e.g. \cite{Ferroni:2011fh}--\cite{GG})\footnote{For a review and an extensive list of references, see \cite{Florkowski:2010zz}.}.  In the most popular applications, such models explain the relative abundances of hadrons, i.e., the ratios of hadron multiplicities. Thermal models can be also  used to analyze the hadron  transverse-momentum spectra and correlations. In the latter case, we often refer to thermal models  as to the {\it hydro-inspired models}. This name reflects the fact that such models do not include the full hydrodynamic evolution but use various hydrodynamics-motivated assumptions about the state of matter at the thermal (kinetic) freeze-out. One of the most popular hydro-inspired models is the {\it blast wave model} originally introduced in \cite{Siemens:1978pb} and adapted to ultra-relativistic energies in \cite{Schnedermann:1993ws}, see also \cite{Florkowski:2004tn} and \cite{Retiere:2003kf}.

With growing beam energies, such as those presently available at the LHC, the final state hadron multiplicities also grow substantially, and hydrodynamic  features of hadron production are expected to appear even in more elementary hadron+hadron and hadron+nucleus collisions, e.g., see Ref.~\cite{Bozek:2013ska}. Quite recently, the blast wave model has been used in this context to analyze high-multiplicity $pp$ collisions at the LHC \cite{Ghosh:2014eqa}. The authors of \cite{Ghosh:2014eqa} found indications of strong transverse radial flow in such events. 

In the present paper, using as the starting point the blast wave model\footnote{Our use of the blast wave model follows similar studies done earlier in the case of heavy-ion collisions \cite{Retiere:2003kf,Kisiel:2006is}.}  featuring a boost-invariant, azimuthally symmetric  fluid expanding in the transverse direction according to the Hubble law \cite{Chojnacki:2004ec}, we show how to  include the possibility of the formation of  thermal clusters as an intermediate step between freeze-out and  particle emission. We are interested, in particular, in the consequences the production of such clusters may have on the measurements of the Bose-Einstein correlations (for a recent review, see \cite{Lisa:2005dd}). 

We note that similar studies have been performed earlier in Refs.~\cite{
Torrieri:2007fb,Zhang:2007wg}. The approach presented in \cite{Torrieri:2007fb} is based on the assumption that the distribution of the particles emitted from a cluster is a gaussian. Within our framework, the particle distribution within a cluster may be arbitrary and the HBT radii are expressed by the moments of the distribution. Moreover, the distribution of clusters assumed in our paper is different from that
proposed in \cite{Torrieri:2007fb}. Our approach differs from that presented in Ref.~\cite{Zhang:2007wg} since we are using a different physical picture. The authors of Ref.~\cite{Zhang:2007wg} assume that the space-time evolution of each cluster/droplet is described by the hydrodynamic equations and the whole system consists of a set of such small hydrodynamic subsystems. In our approach, we model a physical process where a single and large hydrodynamic system breaks first into clusters and later into observed particles (pions). 

  A thermal cluster is characterized by the Boltzmann distribution of the momenta of its decay products:
\ba
   e^{-\beta E^*}=e^{-\beta p^\mu u_\mu},
\ea
where $E^*$ is the energy of the emitted particle in the cluster rest frame, $p^\mu$ is its four-momentum, and $u^\mu$ is the cluster four-velocity. $T=1/\beta$ is the temperature of the cluster.
The new point which we explicitly include in our analysis is the natural condition that the cluster is limited in space-time. This means that in the cluster rest frame  the emission points of its  decay products are distributed in the region described by a positive function $g(x^*)=g(t^*,x^*,y^*,z^*)$ normalized to unity
\ba
\int d^4x^* g(t^*,x^*,y^*,z^*)=1, \lb{gnorm}
\ea 
where $[x^*,y^*,z^*]$ represent the distance from the  center of the cluster to the particle emission point and $t^*$ is the time elapsing from the moment the cluster appears in the system till the particle emission time.  Our aim is to investigate how the finite size of the cluster influences the results and the interpretation of the HBT measurements.

In the next section we define the model by introducing the source function embodying  the formation and decay of clusters. It is based on  the generalized Cooper-Frye formula and the Hubble-like expansion of the fluid. In Section 3 the momentum distribution of particles is evaluated. The HBT correlation functions are discussed in Sections 4 and 5, and in  Section 6 the general formulae for the  HBT radii are given. The results are summarized in the last section. Several Appendices display some details of the algebra needed to obtain the results presented in the paper.

\section{The generalized Cooper-Frye formula}
\label{sect:genCF}

\subsection{Source function}
\label{sect:sourcefun}

Our approach is based on the Cooper-Frye formula \cite{CF}, generalized to the case where matter created at an intermediate stage of the collision process consists of thermal clusters. The starting point is the following expression for the source/emission function
\begin{eqnarray}
S(x,p) = \int d\Sigma_\mu(x_c) \, p^\mu f(x_c)
\int d^4x^* g(x^*)   
  \delta^{(4)}\left(x-x_c- L_c \;x^*\right) e^{-\beta p^\mu u_\mu(x_c)} . \nonumber \\
\label{Sxp1}
\end{eqnarray}
Here $x$ and $p$ are the spacetime position and four-momentum of the emitted particle, $x_c$ and $u^\mu(x_c)$ are the spacetime position and the four-velocity of a cluster, $L_c$ is the Lorentz transformation leading from the cluster rest frame to the  frame where the measurements of the BE correlations are performed and which we shall call "the HBT frame". Finally, $d\Sigma_\mu(x_c)$ is an element of the freeze-out hypersurface which we take in the form
\begin{eqnarray}
d\Sigma_\mu(x_c) &=& S_0  \sigma_\mu(x_c)
\, \delta(\tau_f - \tau_c) d^4x_c =S_0  \sigma_\mu(x_c)
\, \delta(\tau_f - \tau_c)\tau_c d\tau_c d\e_c d^2r_c, \nonumber \\
\label{dSigma}
\end{eqnarray}
where $S_0$ is a normalization constant and the variables $\tau_c$ and $\eta_c$ are the longitudinal proper time and the space-time rapidity of the cluster
\begin{eqnarray}
t_c=\tau_c\cosh\e_c,\sp z_c=\tau_c \sinh\e_c.
\end{eqnarray}
In a similar way, we define the cluster radial distance  from the collision axis and the azimuthal angle in the transverse plane
\begin{eqnarray}
x_c=r_c\cos\phi_c,\sp y_c=r_c\sin\phi_c.
\end{eqnarray}
The four-vector $\sigma^\mu_c=\sigma^\mu(x_c)$ defines the space-time orientation of an element of the freeze-out hypersurface 
\begin{eqnarray}
\sigma^\mu_c = \left(\cosh\eta_c,0,0,\sinh\eta_c\right).
\label{sigma}
\end{eqnarray}

The function $f(x_c)$ in (\ref{Sxp1}) describes the distribution of clusters in space, while the function $g(x^*)$ defines the distribution of the particle emission points in the cluster (in the cluster rest frame). The properties of the functions $f(x_c)$ and $g(x^*)$ will be discussed in more detail below. Here we only note  that for small clusters, i.e., for $x^* \rr  0$, the source function (\ref{Sxp1}) is reduced to the standard emission function \cite{Schnedermann:1993ws}
\begin{eqnarray}
S(x_,p) = \int d\Sigma_\mu(x') p^\mu
\delta^{(4)}(x'-x) \exp(-\beta p^\mu u_\mu(x')) f(x').
\label{Sxpstand}
\end{eqnarray}

Equations (\ref{Sxp1}) and (\ref{dSigma}) allow  to introduce a compact representation of the source function, which highlights its physical interpretation, namely\footnote{From now on we shall omit all constant factors in the source function, since its normalization is irrelevant  for the problems we are discussing in this paper.}
\begin{eqnarray}
S(x,p) = \int d^4x_c  \,\,
 S_c(x_c,u_c)
S_\pi(x_c,u_c,x,p), \label{Sxp2}
\end{eqnarray}
where
\begin{eqnarray}
S_c(x_c, u_c) =   
\, \delta(\tau_f - \tau_c) f(x_c)  \label{Sc}
\end{eqnarray}
and
\begin{eqnarray}
S_\pi(x_c, u_c,x,p) &=& \int d^4x^* \sigma_\mu(x_c) p^\mu  e^{-\beta p_\mu u_c^\mu}  \delta^{(4)}\left(x-x_c- L_c\, x^*\right) g(x^*).
\nonumber \\ \label{Spi}
\end{eqnarray}
Function $S_c(x_c,u_c)$ is the distribution of the cluster four-velocity $u_c$ and space-time position $x_c$, while $S_\pi(x_c,u_c,x,p)$ is the distribution of the  final particles emerging from the cluster decay. Equation (\ref{Sxp2}) shows that the source function can be represented as an integrated product of these two distributions.

\medskip
We assume that  function $f(x_c)$, defining the distribution of clusters in space, depends only on the transverse distance $r_c$. Hence, using Eqs.~(\ref{Sxp2}) and (\ref{Sc}), the source function may be rewritten as
\begin{eqnarray}
S(x,p)= \int r_c dr_c f(r_c) \int d\e_c \int d\phi_c \, S_\pi\left(x_c,u_c,x,p\right).
\label{Sxp3}
\end{eqnarray}

\subsection{Transverse Hubble expansion}
\label{sect:hub}

Since the system is boost-invariant and cylindrically symmetric, the four-velocity of a cluster, $u_c=u(x_c)$, has the form \cite{Florkowski:2010zz}
\begin{eqnarray}
u_c &=&   
\left(\cosh\eta_c \cosh\th_c, 
\sinh\theta_c \cos\phi_c, 
\sinh\theta_c \sin\phi_c, 
\sinh\eta_c \cosh\theta_c \right).
\label{u}
\end{eqnarray}
In addition, we assume that the transverse rapidity of the cluster $\theta_c$ and its position $r_c$ are related by the condition of the radial Hubble-like  flow \cite{Chojnacki:2004ec}. This leads to the expressions
\begin{eqnarray}
\sinh\theta_c = \omega r_c, \quad
\cosh\theta_c = \sqrt{1+\omega^2 r_c^2},
\label{hub}
\end{eqnarray} 
where $\omega$ is the parameter controlling the magnitude of the transverse flow.

The particle four-momentum is parameterized in the standard way in terms of rapidity, $y$, transverse momentum, $p_\perp$, transverse mass, $m_\perp$, and the azimuthal angle in the transverse plane, $\phi_p$,
\begin{eqnarray}
p = \left(m_\perp \cosh y, p_\perp \cos\phi_p, p_\perp \sin\phi_p, m_\perp \sinh y \right).
\label{p}
\end{eqnarray}
Then, the scalar product of $p$ and $u_c$ is
\begin{eqnarray}
p \cdot u_c = m_\perp \cosh(y-\eta_c) \cosh\theta_c - p_\perp \cos(\phi_p-\phi_c) \sinh\theta_c.
\label{pu}
\end{eqnarray}
This form is used in the thermal Boltzmann distribution. In a similar way we obtain the factor $p \cdot \sigma_c$ needed to define the element of the freeze-out hypersurface\footnote{The form of (\ref{psigma}) follows directly from (\ref{p}) and (\ref{pu}). Other forms are also possible here if one assumes different freeze-out conditions. Using (\ref{dSigma}) and (\ref{sigma}) we follow the most popular version of the blast wave model. }
\begin{eqnarray}
p \cdot \sigma_c = m_\perp  \cosh(y-\eta_c).
\label{psigma}
\end{eqnarray}

%%%%%%%%%%%%%%%%%%%%%%%%%%%%%%%%%%%%%%%%%%%
\subsection{Distribution of the emitted particles in a thermal cluster}
\label{sect:decay}
%%%%%%%%%%%%%%%%%%%%%%%%%%%%%%%%%%%%%%%%%%%%%%%%

The decay distribution can be written as 
\begin{eqnarray}
S_\pi(x_c,u_c,x,p) = p_\mu \sigma^\mu_c
\exp(-\beta p_\mu u_c^\mu) S^*(x_c,x,u_c).
\label{Spi2}
\end{eqnarray}
The first two factors in (\ref{Spi2}) describe the momentum distribution. We have  
\begin{eqnarray}
p_\mu \sigma^\mu_c 
\exp(-\beta p_\mu u_c^\mu) &=& m_\p \cosh(y-\eta_c)
\label{momdistr} \\
&& \hspace{-2cm} \times \exp\left[-
\beta m_\p \cosh \theta_c \cosh (\eta_c-y) +\beta p_\p \sinh\theta_c \cos \phi \right], \nonumber
\end{eqnarray}
where $\phi=\phi_c-\phi_p$ is the angle in the transverse plane between $\vec{u}_{c,\perp}$ and $\vec{p}_\p$. The last factor in (\ref{Spi2}), i.e. the function $S^*(x_c,x,u_c)$, describes the distribution of the points of  particle emission  from the cluster, which is discussed in greater detail in Appendix \ref{appL},  
\begin{eqnarray}
S^*(x_c,x,u_c)&=&\int d^4 x^*g(x^*)
\delta(t-t_c-T) \delta (x-x_c-X) \nonumber \\
&& \times \delta(y-y_c-Y)\delta(z-z_c-Z),
\lb{star}
\end{eqnarray}
where
\begin{eqnarray}
T &=& \cosh\eta_c \left( t^* \cosh\theta_c + 
x^* \sinh\theta_c \right)+z^*\sinh\e_c,
\nonumber \\
X &=& x^* \cos\phi_c \cosh\theta_c - y^* \sin\phi_c + t^* \cos\phi_c \sinh\theta_c, 
\nonumber \\
Y &=& y^* \cos\phi_c +x^* \cosh\theta_c \sin\phi_c + t^* \sin\phi_c \sinh\theta_c \nonumber \\
Z &=& \sinh\eta_c \left(t^* \cosh\theta_c + 
x^* \sinh\theta_c \right)+z^*\cosh\e_c .
\label{xmu}
\end{eqnarray}
Integration over $d^4x^*$ is easy and gives
\ba
S^*= g(\hat{t},\hat{x},\hat{y},\hat{z})
\ea
with
\begin{eqnarray}
\hat{t} &=& (T'\cosh\e_c-Z'\sinh\e_c)\cosh\th_c-(Y'\sin\phi_c+X'\cos\phi_c)
\sinh\th_c, \nonumber \\
\hat{x} &=& -(T'\cosh\e_c-Z'\sinh\e_c)\sinh\th_c+(Y'\sin\phi_c+X'\cos\phi_c)
\cosh\th_c, \nonumber \\
\hat{y} &=& Y'\cos\phi_c-X'\sin\phi_c,
\nonumber \\
\hat{z} &=& -T'\sinh\e_c+Z'\cosh\e_c,
\lb{starhat}
\end{eqnarray}
and $X^{'\mu}\equiv [T',X',Y',Z']= (x-x_c)^\mu$.

%%%%%%%%%%%%%%%%%%%%%%%%%%%%%%%%%%%%%%%%%%%
%%%%%%%%%%%%%%%%%%%%%%%%%%%%%%%%%%%%%%%%%%% 
\section {Momentum distribution}
\label{sect:momdis}
%%%%%%%%%%%%%%%%%%%%%%%%%%%%%%%%%%%%%%%%%%%
%%%%%%%%%%%%%%%%%%%%%%%%%%%%%%%%%%%%%%%%%%% 

By definition, the integral of the source function $S(x,p)$ over the spacetime coordinates gives the momentum distribution
\ba
\frac{dN}{dy d^2p_\perp} = W(p)= \int d^4x \,S(p,x).
\ea
The explicit calculation starting from Eq.~(\ref{Sxp1}) yields
\begin{eqnarray}
W(p) &=& \int d^4x^*  \int d\Sigma_\mu(x_c) p^\mu
 e^{-\beta p^\mu u_\mu(x_c)} f(x_c)  g(x^*) \nonumber \\
 &=& \int d\Sigma_\mu(x_c) p^\mu e^{-\beta p^\mu u_\mu(x_c)} f(x_c) .
 \label{dndyd2pt}
\end{eqnarray}
Thus, the particle momentum distribution is given by the same expression as that used in the standard Cooper-Frye formula (with the particle space-time coordinates replaced by the cluster coordinates).  The integration over $d^4x$ cancels the four Dirac delta functions appearing in (\ref{star}) and leads  to the formula
\begin{eqnarray}
W(p) =  m_\p \int r_c dr_c f(r_c) \int d\eta_c \int d\phi_c   \cosh(\eta_c-y)
 e^{-U \cosh (\eta_c-y) + V \cos \phi_c }, 
 \nonumber \\ \lb{wint}
\end{eqnarray}
with
\ba
U=\beta m_\p \cosh\theta_c, \sp
V=\beta p_\p \sinh\theta_c.
\label{UandV}
\ea
Integration over $\eta_c$ and $\phi_c$ gives
\ba
W(p_\p)=  m_\p \int r_c dr_c  f(r_c)   K_1(U)) I_0(V), 
\ea
which agrees with a formula commonly used to interpret the transverse-momentum spectra \cite{Schnedermann:1993ws}.

%%%%%%%%%%%%%%%%%%%%%%%%%%%%%%%%%%%%%%%%%%%
%%%%%%%%%%%%%%%%%%%%%%%%%%%%%%%%%%%%%%%%%%% 
\section{HBT correlation function}
%%%%%%%%%%%%%%%%%%%%%%%%%%%%%%%%%%%%%%%%%%%
%%%%%%%%%%%%%%%%%%%%%%%%%%%%%%%%%%%%%%%%%%% 

Assuming that one can neglect correlations between the produced particles\footnote{Although, as pointed out in \cite{Bialas:2013oza}, this assumption may distort significantly the results for $Q$ exceeding the inverse size of the system, it is not  restrictive at small $Q$, the region which is of interest in this paper.},  the distribution of two identical bosons can be expressed in terms of the Fourier transform of the source function
\ba
W(p_1,p_2)=W(p_1)W(p_2) +|H(P,Q)|^2
\ea
with
\begin{eqnarray} 
H(P,Q)&=&\int d^4x e^{iQ \cdot x} S(x,P).
\label{hpq}
\end{eqnarray}
Here $Q=p_1-p_2$ and $\vec{P}=(\vec{p}_1+\vec{p}_2)/2$. The time-component of the four-vector $P$ is not uniquely defined. We shall adopt the convention $P_0=\sqrt{m^2+|\vec{P}|^2}$ \cite{Pratt:2008qv}. In Appendix E we discuss the consequences of another  relation, $P_0=(p_{01}+p_{02})/2$ \cite{Lisa:2005dd}.  

The source function $S(x,P)$ appearing in (\ref{hpq}) is given by  our initial definition, see Eq.~(\ref{Sxp1}), with $p$ replaced by $P$, namely  
\begin{eqnarray}
S(x,P)
&=& \int d^4x^*g(x^*) 
\int d\Sigma_\mu(x_c) P^\mu
\exp\left(-\beta P^\mu u_\mu(x_c)\right) f(r_c) 
\nonumber \\
&& \times 
\delta\left(t-t_c-T\right) 
\delta(z-z_c-Z)\delta\left(x- x_c - X \right)
\delta\left(y - y_c - Y \right) 
\nonumber \\
&=& \int d\Sigma_\mu(x_c) P^\mu 
\exp\left(-\beta P^\mu u_\mu(x_c)\right) f(r_c)
S^*(x_c,x,u_c).
\label{SxP}
\end{eqnarray}
In the last line in (\ref{SxP}) we used our definition of the function $S^*(x_c,x,u_c)$, see Eq.~(\ref{star}).

Equations (\ref{hpq}) and (\ref{SxP}) allow us to write the compact expression for the Fourier transform of the source function
\begin{eqnarray}
\hspace{-0.75cm}
H(P,Q) &=& \int d\Sigma_\mu(x_c) P^\mu 
 \exp\left(-\beta P^\mu  u_\mu(x_c)\right) f(r_c)
e^{i Q \cdot x_c}
{\cal G}(x_c,Q)
\label{hpq1}
\end{eqnarray}
where
\begin{eqnarray}
{\cal G}(x_c,Q) = \int d^4x^*\exp\left[i (Q \cdot X) \right] g(x^*)
\label{calG}
\end{eqnarray}
with $X^\mu$  given by Eq.~(\ref{xmu}).

\section {Kinematics of the Fourier transform}

We shall  work in the so-called LCMS system (i.e. our HBT system is the LCMS system)  in which $P_z=0$, i.e. $p_{1z}=-p_{2z}$ and $y_{\rm pair}=0$. In this frame the substitution $p\rr P$ is simply realized by the change $m_\p\rr \sqrt{P_0^2-P_z^2}= P_0$. Starting directly from (\ref{hpq}) and (\ref{SxP}) we find
\begin{eqnarray}
H(P,Q) &=& P_0 \int d^4x^*  
\int r_c\, dr_c f(r_c) \int d\phi_c \nonumber \\
&& \times \int d\eta_c  \cosh\eta_c e^{-U \cosh\eta_c + V \cos\phi_c - i \Phi }g(x^*)
\label{FT1}
\end{eqnarray}
where now $U=\beta P_0\cosh\theta_c$, $V=\beta P_\perp \sinh\theta_c$, and the phase $\Phi$ is given
by the formula
\begin{eqnarray}
\Phi = -Q_0 (t_c+T) +Q_z (z_c+Z) + Q_x (x_c + X) + Q_y (y_c +Y). \label{Phi}
\end{eqnarray}
The phase $\Phi$ depends on the  relative direction of $\vec{P}=(P_\perp,0,0)$ and $\vec{Q}$. One considers three regimes:  

\begin{itemize}
\item[i)] {\it long} direction: $Q_x = Q_y=0$, $Q_z=q $, and $Q_0=0$,
\begin{eqnarray}
\Phi_{\rm long} &=& q  \chi_{\rm long}   ,\;\;\;\;
\chi_{\rm long} =  \tau_f \sinh\eta_c+Z; \label{chil}  
\end{eqnarray}
\item[ii)] {\it side}  direction: $Q_x = Q_z=0$, $Q_y = q$, and $Q_0=0$,
\begin{eqnarray}
\Phi_{\rm side} &=& q \, \chi_{\rm side}, \label{chis} \;\;\;\;
\chi_{\rm side} = r_c \sin\phi_c  +Y;
\end{eqnarray}
\item[iii)] {\it out} direction: $Q_y = Q_z=0$, $Q_x = q$ and
\begin{eqnarray}
Q_0 = \sqrt{m^2 + (P_\perp+q/2)^2} - \sqrt{m^2 + (P_\perp-q/2)^2},
\label{Q0}
\end{eqnarray}
\begin{eqnarray}
\!\!\!\!\Phi_{\rm out} &=& q \, \chi_{\rm out}, \label{chio} \;\;\;
\chi_{\rm out} = 
-\frac{Q_0}{q}[ \t \cosh\e_c +T] + r_c \cos\phi_c +X.
%\nonumber
\end{eqnarray}
\end{itemize}
For  small $q$, which is sufficient to obtain the HBT radii (as described in more detail below), we find $Q_0/q \approx  P_\perp /\sqrt{m^2+P_\perp^2}\equiv \zeta$. For arbitrary values of $q$ one should use explicitly  formula (\ref{Q0}).   

With the help of the notation introduced above the three desired versions of the Fourier transform may be written  as one universal formula
\begin{eqnarray}
H_d(P_\perp,q) = P_0 \int d^4x^* g(x^*)
\int r_c\, dr_c f(r_c) \int d\phi_c e^{ V \cos\phi_c }\nn
 \times \int d\eta_c  \cosh\eta_c e^{-U \cosh\eta_c}e^{ - i q \chi_d} \equiv H(P_\perp,q=0)\left<e^{ - i q \chi_d}\right> .
\label{FT2}
\end{eqnarray}
The subscript $d$ stands for ``long'', ``side'', and ``out''.  

If clusters are not present, $H_d(P_\perp,q) $ can be explicitly expressed in terms of integrals of  Bessel functions. The corresponding formulae are given in Appendix D.

Equation (\ref{FT2}) is the basis for the consideration of the ``numerator'' contributions to the HBT radii discussed in Appendix \ref{appH}. The complementary ``denominator'' contributions are discussed in Appendix \ref{appW}. Let us note that sometimes it is assumed that the denominator 
does not contribute to the HBT radii \cite{Lisa:2005dd}.

\section {The HBT radii}

Experiments usually measure the correlation function defined as
\ba
C(p_1,p_2)\equiv \frac{W(p_1,p_2)}{W(p_1)W(p_2)} -1=\frac{|H(P,Q)|^2}{W(p_1)W(p_2)}
\ea
The measured HBT radii are obtained from the fit to the correlation function in the gaussian form separately for each of the directions $long$, $side$ and $out$
\ba
C(p_1,p_2)=e^{-R^2_{HBT}q^2}
\ea
with $q$ given by (\ref{chis}). This means that they can be evaluated as the logarithmic derivative at $q=0$
\ba
R^2_{HBT}=-\frac{d \log[C(p_1,p_2)]}{dq^2} \equiv R^2_H-R^2_W
\ea
with
\ba
R^2_H=-\left\{\frac{dH(P,q)/dq^2}{H(P,q)}+\frac{dH^*(P,q)/dq^2}{H^*(P,q)}\right\}_{q=0}, \nonumber \\
R^2_W=-\left\{\frac{dW(p_1)/dq^2}{W(p_1)}+\frac{dW(p_2)/dq^2}{W(p_2)}\right\}_{q=0}.
\ea

Using the formulae of the previous section it is thus possible to derive the expressions 
for the HBT radii for all the three configurations. In this section we only give the final results. The algebra is outlined in Appendices \ref{appH} and \ref{appW}.

In Appendix \ref{appH} it is shown that 
\ba
R^2_{Hd}= \left<\chi_d^2\right>-\left<\chi_d\right>^2
=\left<\left(\chi_d-\left<\chi_d\right>\right)^2
\right>,
\ea
where the average $\left<\O\right>$ is defined as
\ba
\left<\O\right>=\frac{\int r_cdr_c f(r_c)d\phi_c d\e_c \cosh\e_c e^{-U\cosh\e_c} e^{V\cos\phi_c} \int d^4x^*g(x^*) \O}{\int r_cdr_c f(r_c)d\phi_c d\e_c \cosh\e_c e^{-U\cosh\e_c} e^{V\cos\phi_c} \int d^4x^*g(x^*)}.
\ea

These averages can be expressed in terms of  integrals of  Bessel functions as shown in Appendix D. The results are listed below.
Denoting
\ba
 \rho^2\equiv \left<x^{*2}\right>=\left<y^{*2}\right>;\;\;\rho_z^2=\left<z^{*2}\right>;\;\;\;\rho_t^2=\left<t^{*2}\right>;\;\;\d_t=\left<t^*\right>; \lb{skr}
\ea
where $\langle \O^* \rangle \equiv \int d^4x^* \O \, g(x^*) $, one obtains
\begin{eqnarray}
\left<\chi_{long}\right> &=& \left<\chi_{side}\right>=0, 
\nonumber \\
\left<\chi_{out}\right> &=& \frac{\int r_cdr_cf(r_c)\l_1 I_1(V)K_1(U)}{\int r_cdr_cf(r_c)I_0(V)K_1(U)} -
\frac{\int r_cdr_cf(r_c) \l_2I_0(V)K_0''(U)}{\int r_cdr_cf(r_c)I_0(V)K_1(U)},
\nonumber \\
\end{eqnarray}
where
\begin{equation}
\l_1 = r_c+\d_t\sinh\th_c;\sp \l_2=\zeta(\t+\d_t\cosh\th_c).
\end{equation}
The average $\left<\chi_{long}^2\right>$ is given by the formula
\ba
\left<\chi_{long}^2\right>=\frac{\int r_cdr_cf(r_c)\l_2I_0(V)\left[\kappa_l[K_1''(U)-K_1(U)]+\rho_z^2K_1''(U)\right] }{\int r_cdr_cf(r_c)I_0(V)K_1(U)},
\ea
with 
\ba
\kappa_l=\t^2+\rho_t^2\cosh^2\th_c+\rho^2\sinh^2\th_c +2\t\d_t\cosh\th_c.  \lb{kl}
\ea
For the {\it side} direction the result is
\ba
\left<\chi_{side}^2\right>=
\frac{\int r_cdr_cf(r_c)\left[\kappa_s [I_0(V)-I_1'(V)]+\rho^2I_1'(V)\right] K_1(U)}{\int r_cdr_cf(r_c)I_0(V)K_1(U)}
\ea
with 
\ba
\kappa_s=r_c^2+\rho^2\cosh^2\th_c
+\rho_t^2\sinh^2\th_c+2r_c\d_t\sinh\th_c,  \lb{ks}
\ea
while for the {\it out} direction we find
\begin{eqnarray}
\left<\chi_{out}^2\right> &=& \zeta^2
\frac{\int r_cdr_cf(r_c) I_0(V)\left[\kappa_1K_1''(U)+\rho_z^2[K_1''(U)-K_1(U)]\right]}{\int r_cdr_cf(r_c)I_0(V)K_1(U)}
\nonumber \\
&&-2\zeta\frac{\int r_cdr_cf(r_c)\kappa_2 I_1(V)K_0''(U)}{\int r_cdr_cf(r_c)I_0(V)K_1(U)}
\nonumber \\
&& + \frac{\int r_cdr_cf(r_c) [\kappa_3 I_0''(V)+\rho^2[I_0(V)-I_0''(V)]]K_1(U)}{\int r_cdr_cf(r_c)I_0(V)K_1(U)}
\end{eqnarray}

with
\begin{eqnarray}
\kappa_1 &=& \t^2+\rho_t^2\cosh^2\th_c+\rho^2\sinh^2\th_c
+2\t\d_t\cosh\th_c, \nonumber \\
\kappa_2 &=& \t( r_c+\d_t\sinh\th_c)+r_c\d_t\cosh\th_c
+[\rho^2+\rho_t^2]\sinh\th_c\cosh\th_c,
\nonumber \\
\kappa_3 &=& r_c^2+\rho^2\cosh^2\th_c
+\rho_t^2\sinh^2\th_c+2r_c\d_t\sinh\th_c. \lb{k123}
\end{eqnarray}

Evaluation of $R_W^2$ is given in Appendix C. The results for different directions are as follows

\ba
R^2_{W, \rm long}=0, \label{R2Wlong}
\ea
\begin{eqnarray}
&& R_{W, \rm side}^2 = -\frac1{4M_\perp^2} \label{R2Wside} \\
&&-\frac{\beta^2}4
\frac{\int dr_c^2 f(r_c)\left[\sinh^2\theta_c \, (I_0' (V)/V) K_1(U)+\cosh^2\theta_c I_0(V)K_1'(U)/U\right]}{\int dr_c^2 f(r_c) I_0(V)K_1(U)}, \nonumber
\end{eqnarray}

\begin{eqnarray}
&& R^2_{W, \rm out}=\frac{P_\p^2-m^2}{4M_\perp^4} \label{R2Wout} \\
&&
+\frac{\beta^2}4\left[\frac{\int dr_c^2 f(r_c)\left[\sinh\theta_c I_0'(V)K_1(U)+\cosh\theta_c I_0(V)K_1'(U)\zeta\right]}{\int dr_c^2 f(r_c) I_0(V)K_1(U)}\right]^2 \nonumber \\
&&-
\frac{\beta^2}4 \frac{\int dr_c^2 f(r_c)[\sinh^2\theta_c I_0''(V)K_1(U)+2\sinh\theta_c\cosh\theta_c I_0'(V)K_1'(U)\zeta]}{\int dr_c^2 f(r_c)I_0(V)K_1(U)}
\nonumber \\
&&-
\frac{\beta^2}4 \frac{\int dr_c^2 f(r_c) \cosh^2\theta_c I_0(V)K_1''(U)\zeta^2 }{\int dr_c^2 f(r_c)I_0(V)K_1(U)}
\nonumber \\
&& -
\frac{\beta^2 m^2}{4M_\perp^2} \frac{\int dr_c^2 f(r_c)\cosh^2\theta_c I_0(V)K_1'(U)/U}{\int dr_c^2 f(r_c)I_0(V)K_1(U)}, \nonumber
\end{eqnarray}
where  $\zeta=P_\p/M_\perp$. 

For the reader's convenience we also include below all the needed relations for the Bessel functions:
\begin{eqnarray}
K_1'(a) &=& -K_0(a)-K_1(a)/a, \nonumber \\
K_1''(a)&=& K_0(a)/a +K_1(a)+2K_1(a)/a^2 
\nonumber \\
K_2(a) &=& K_0(a)+2K_1(a)/a, \nonumber \\ I_0'(a) &=& I_1(a),
\nonumber \\
I_0''(a) &=& I_1'(a)=I_0(a)-I_1(a)/a. 
\end{eqnarray}

\section{Summary}

The observed  success of the statistical model in explaining many features of particle production processes in  high-energy collisions suggests that particles are produced in form of "thermal clusters" which decay into the observed final state. In the present  paper we discussed how this mechanism can influence  measurements of  quantum interference. To this end we have generalized the well-known blast wave model \cite{Schnedermann:1993ws} to include the production of  thermal clusters. 
The novel element of our approach is  introducing  the final size and life-time of a cluster  which, as one may expect,  modifies the  interpretation of the HBT measurements and makes the model more flexible. The explicit formulae for the correlation functions and for the HBT radii have been derived in a form which is ready for direct application.

 As the presence of thermal clusters is  an almost unavoidable consequence of the success of the statistical model of particle production, we feel that our work provides the necessary tools which may serve to verify the statistical picture on a more fundamental level.

 Furthermore, determination of the cluster parameters and verification if they reveal some universal features may be an important contribution to understanding of the statistical model.

In conclusion, we have shown that the presence of the thermal clusters does not invalidate  the significance  of the measurements of quantum interference but, on the contrary,  allows to extract from them even more interesting information.

\vspace{0.3cm}
 
Acknowledgments:  This investigation was supported in part by the NCN Grants UMO-2013/09/B/ST2/00497 and DEC-2012/06/A/ST2/00390.

\appendix

\section{Lorentz transformation connecting the cluster's rest frame and the HBT frame.}
\label{appL}

The active Lorentz transformation $L_c$ leading from the cluster local rest frame (CRF), where its velocity is \mbox{$u^*=(1,0,0,0)$}, to the HBT frame, where the velocity is $u_c$, may be represented as a composition of three Lorentz transformations: a Lorentz boost along the $x$-axis,
\begin{equation}
 L_{(x)}(\theta_c) = \left(
\begin{array}{cccc}
\cosh\theta_c & \sinh\theta_c & 0 & 0 \\
\sinh\theta_c & \cosh\theta_c & 0 & 0 \\
0 & 0 & 1 & 0 \\
0 & 0 & 0 & 1
\end{array} \right),
\label{Lx}
\end{equation}
a rotation around the $z$-axis,
\begin{equation}
 R_{(xy)}(\phi_c) = \left(
\begin{array}{cccc}
1 & 0 & 0 & 0 \\
0 & \cos\phi_c & -\sin\phi_c & 0 \\
0 & \sin\phi_c & \cos\phi_c & 0 \\
0 & 0 & 0 & 1
\end{array} \right) , 
\label{Rxy}
\end{equation}
and a boost along the $z$-axis,
\begin{equation}
 L_{(z)}(\eta_c) = \left(
\begin{array}{cccc}
\cosh\eta_c & 0 & 0 & \sinh\eta_c \\
0 & 1 & 0 & 0 \\
0 & 0 & 1 & 0 \\
\sinh\eta_c & 0 & 0 & \cosh\eta_c
\end{array} \right).
\label{Lz}
\end{equation}
Indeed, by direct multiplication of the matrices one can check that
\begin{eqnarray}
u_c &=& L_c \, u^* =  L_{(z)}(\eta_c) R_{(xy)}(\phi_c)  L_{(x)}(\theta_c) u^*.
\end{eqnarray}
In order to change from the HBT frame to CRF we perform simply the inverse transformation
\begin{eqnarray}
u^* = L_c^{-1} u_c &=& L_{(x)}^{-1}(\theta_c) R_{(xy)}^{-1}(\phi_c)   L_{(z)}^{-1}(\eta_c) u_c \nonumber \\
&=& L_{(x)}(-\theta_c) R_{(xy)}(-\phi_c)   L_{(z)}(-\eta_c) u_c.
\label{ustar} 
\end{eqnarray}

In the HBT frame  the fluid element with four-velocity $u_c$ is placed at the space-time point $x_c$ with the coordinates
\begin{eqnarray}
x_c = (\tau \cosh\eta_c, r_c \cos\phi_c, r_c \sinh\phi_c, \tau \sinh\eta_c),
\label{xc}
\end{eqnarray} 
and a particle is emitted from the space-time point
\begin{eqnarray}
x = (t,x,y,z).
\label{xp}
\end{eqnarray}
Then, the Lorentz transformation of the coordinate difference $x^*\equiv [t^*,x^*,y^*,z^*]$ is
\begin{eqnarray}
\!\!x^\mu\!-\!x_c^\mu &=& L_c  \, x^* \! \nonumber
\\
&=&\!\left(
\begin{array}{c}
\cosh\eta_c \left( t^* \cosh\theta_c + 
x^* \sinh\theta_c \right)+z^*\sinh\e_c \\
x^* \cos\phi_c \cosh\theta_c - y^* \sin\phi_c + t^* \cos\phi_c \sinh\theta_c \\
y^* \cos\phi_c +x^* \cosh\theta_c \sin\phi_c + t^* \sin\phi_c \sinh\theta_c \\
\sinh\eta_c \left(t^* \cosh\theta_c + 
x^* \sinh\theta_c \right)+z^*\cosh\e_c \\
\end{array} \right).
\label{xcmx}
\end{eqnarray}
%

%%%%%%%%%%%%%%%%%%%%%%%%%%%%%%%%%%%%%%%%%%%%%%%%
\section{HBT radii --- numerator contributions}
\label{appH}
%%%%%%%%%%%%%%%%%%%%%%%%%%%%%%%%%%%%%%%%%%%%%%%%

As demonstrated by Eq.~(\ref{FT2}), for small values of the momentum difference $Q$ the Fourier transform appearing in the numerator of the correlation function can be schematically written as
\ba
H=\int d\Omega s(\Omega,P)  e^{iQ \cdot x},
\ea
where $\Omega=[r_c,\phi_c,\e_c ; x^*]$  denotes, symbolically, all variables to be integrated over, $d\Omega= r_c f(r_c)dr_c d\phi_c d\e_c \cosh\e_c d^4x^* g(x^*)$, and
\ba
s(\Omega,P)= P_0  e^{-U\cosh\e_c} e^{V\cos\phi_c} .
\ea
For the three directions we write $Q\cdot x=q \chi $, where $\chi$ is independent of $q$ and the three relevant  options  for $\chi$ are given by Eqs.~(\ref{chil})--(\ref{chio}).

We need to evaluate the derivative $d\log H/dq^2$ at $q=0$. To this end we observe that, up to the second order in $q$,
\begin{eqnarray}
 \log H &=& \log\left[\int d\Omega s(\Omega,P)  \left(1+iq\chi  -q^2\chi^2/2\right)\right]
\\ &=& \log\left[\int d\Omega s(\Omega,P)\right]+\log\left[1+i q \langle\chi\rangle -q^2\left<\chi^2\right>/2\right], \nonumber
\end{eqnarray}
where
\ba
 <\O>\equiv \frac{\int d\Omega s(\Omega,P)\O}{\int d\Omega s(\Omega,P)}.
\ea
Consequently, one finds
\ba
\hspace{-0.5cm} R^2_H=-\left[\frac{d\log H}{dq^2}+\frac{d\log H^*}{dq^2}\right]_{q=0}=\left<\chi^2\right>-<\chi>^2=
\left<\left(\chi-\langle\chi\rangle\right)^2\right>.
\ea
where the asterisk denotes complex conjugation.
Using these formulae and  the explicit expressions (\ref{chil}), (\ref{chis}), and (\ref{chio}) one can evaluate the radii for all directions. The symmetries of $s(\Omega,P)$ imply 
\begin{eqnarray}
\left<\chi_{long}\right> &=& \left<\chi_{side}\right>=0,
\nonumber \\
\left<\chi_{out}\right> &=& 
\left<(r_c+\d_t\sinh \th_c)\cos\phi_c\right>
-\zeta\left<(\t+\d_t\cosh\th_c)\cosh\e_c\right>.
\end{eqnarray}
Using the abbreviation (\ref{skr})  we have
\ba
\left<\chi_{long}^2\right>=\left<\kappa_l\sinh^2\e_c+\rho_z^2\cosh^2\e_c\right>
\ea
\ba
\left<\chi_{side}^2\right>=\left<\kappa_s\sin^2\phi_c+\rho^2\cos^2\phi_c\right>
\ea
\begin{eqnarray}
\left<\chi_{out}^2\right>
&=&\zeta^2\left<\kappa_1\cosh^2\e_c
+\rho_z^2\sinh^2\e_c
-2\kappa_2\cosh\e_c\cos\phi_c\right>
\nonumber \\
&& +\left<\kappa_3\cos^2\phi_c
+\rho^2\sin^2\phi_c\right>
\end{eqnarray}
with $\kappa_l$,  $\kappa_s$ and  $\kappa_1,\kappa_2,\kappa_3$ given by (\ref{kl}), (\ref{ks})and (\ref{k123}).

  Observing that 
\begin{eqnarray}
\int \frac{d\e}2 \cosh^n\e e^{-U\cosh\e} &=& (-1)^n\frac{d^nK_0(U)}{dU^n},
\nonumber \\
 \int \frac{d\phi}{2\pi} \cos^n\phi e^{V\cos\phi}
 &=&  \frac{d^nI_0(V)}{dV^n}
\end{eqnarray}
one can express all these averages in terms of  Bessel functions. The resulting formulae are listed in  Section 6.

%%%%%%%%%%%%%%%%%%%%%%%%%%%%%%%%%%%%%%%%%%%%%%%%
\section{HBT radii --- denominator contributions}
\label{appW}
%%%%%%%%%%%%%%%%%%%%%%%%%%%%%%%%%%%%%%%%%%%%%%%%  

The contribution of the denominator to the HBT radii is given by the formula
\ba
R^2_W=-\frac{d}{dq_\p^2} \log \left[w(\vec{p}_+)w(\vec{p}_-)\right] \lb{c1}
\ea
with $\vec{p}_\pm=\vec{P}_\p\pm\vec{q}_\p/2$ and
where the derivative is evaluated at $q_\p=0$. Since in the \emph{long} case $q_\p\equiv 0$, we find immediately that $R^2_{W,long} = 0$. For the other two cases the functions $w$ are defined by Eq.~(\ref{wint}). To evaluate (\ref{c1}) one needs  them only up to second order in $q_\p$.

The contributions from the $m_T$ factors are easily evaluated. The results are given as the first terms in (\ref{R2Wside}) and (\ref{R2Wout}). The other contributions are more involved.

Consider first the $side$ direction. In this case we have
\ba
U(\vec{p}_\pm)=U[1+q_\p^2/8m_\p^2];\;\;\;V(\vec{p}_\pm)=V[1+q_\p^2/8P_\p^2];\;\;\;
\ea
where $U$ and $V$ are given by (\ref{UandV}). Expanding $e^{-U(\vec{p}_\pm)\cosh\e+V(\vec{p}_\pm)\cos\phi}$ in powers of $q_\p^2$ and integrating term by term  it is straightforward to  obtain  Eq. (\ref{R2Wside}). 
For the $out$ direction we have
\begin{eqnarray} 
U(\vec{p}_\pm) &=& U(\vec{P})\left[1\pm P_\p q_\p/2m_\p^2+m^2q_\p^2/8m_\p^4\right],
\nonumber \\
V(\vec{p}_\pm) &=& V(\vec{P})[1\pm q_\p/2].
\end{eqnarray}
Consequently, one finds
\begin{eqnarray}
&& e^{-U(\vec{p}_\pm)\cosh\e}
e^{V(\vec{p}_\pm)\cos\phi} =
e^{-U\cosh\e+V\cos\phi}
\nonumber \\ 
&& \times\left[1\mp\frac{\beta\cosh\th q_\p\zeta}2\cosh\e-\frac{\beta\cosh\th q_\p^2m^2}{8m_\p^3}\cosh\e
+\frac{\beta^2\cosh^2\th\zeta^2q_\p^2}8
\cosh^2\e\right]
\nonumber \\
&& 
\times \left[1\pm\frac{\beta\sinh \th q_\p}2\cos\phi +\frac{\beta^2\sinh^2\th q_\p^2}8\cos^2\phi\right].
\end{eqnarray}
Observing that the terms linear in $q_\p$ cancel  when one considers the  logarithm of the product $w(\vec{p}_+)w(\vec{p}_-)$,  one obtains, after some algebra, Eq.~(\ref{R2Wout}).

\vspace{1cm}

%%%%%%%%%%%%%%%%%%%%%%%%%%%%%%%%%%%%%%%%%%
\section{Angle and space-time integrals}
\label{appintegrals}
%%%%%%%%%%%%%%%%%%%%%%%%%%%%%%%%%%%%%%%%%%

In our analysis we frequently have to evaluate  integrals of the form
\begin{eqnarray}
G &\equiv& P_0 \int r_c dr_c f(r_c) \int   d\phi\, d\e  \cosh \e  e^{-U' \cosh \e + i a \sinh\eta} 
e^{V' \cos\phi-i b \sin\phi}. \nonumber \\
\end{eqnarray}
In the case of the {\it long} direction we have: $U'=U$, $V'=V$, $a= q \tau_f$, and $b=0$. Hence we may write
\begin{eqnarray}
G_{long} &=&  P_0\int r_c dr_c f(r_c) \int d\phi e^{V \cos\phi} D_{long}(U,a),
\end{eqnarray}
where
\begin{eqnarray}
D_{long}(U,a) &\equiv& \int  d\e  \cosh \e e^{-U\cosh \e} e^{i a\sinh \e} =
 -\frac{d}{dU}\int d\e e^{-U \cosh \e} e^{i a\sinh \e}. \nonumber
\end{eqnarray}
Since
\ba
\hspace{-0.5cm} \int d\e e^{-U \cosh \e} e^{ia\sinh \e}=\int d\e e^{-\sqrt{U^2+a^2}\cosh(\e-\e')}=2K_0[\sqrt{U^2+a^2}]
\ea
we finally obtain
\ba
G_{long}=4 \pi P_0 \int r_cdr_cf(r_c) I_0(V) UK_1(U_l)/U_l;\;\;\; U_l=\sqrt{U^2+a^2}.
\ea
In the case of the {\it side} direction we have: $U'=U$, $V'=V$, $a=0$, and $b=q r_c$. This leads us to the expression
\begin{eqnarray}
G_{side} &=& P_0 \int r_cdr_cf(r_c) \int d\e \cosh \e e^{-U \cosh \e}D_{side}(V,b),
\end{eqnarray}
where
\begin{eqnarray}
D_{side}(V,b) &\equiv& \int d\phi e^{-ib \sin \phi}e^{V \cos\phi}=\int_0^{2\pi} d\phi 
e^{V_s \cos (\phi-\phi')}= 2\pi I_0(V_s)
\nonumber
\end{eqnarray}
and $V_s=\sqrt{V^2-b^2}$. Thus, finally,
\ba 
G_{side}= 4 \pi P_0 \int r_cdr_cf(r_c) I_0(V_s) K_1(U).
\ea 
If $V^2<Q^2r^2$, $V_s$ is  imaginary and the function $I_0(V_s))$ should  be replaced by $J_0(|V_s|)$. In the case of the {\it out} direction we use: $U'=U-iQ_0 \tau_f$, $V'= V + i q r_c$, $a=b=0$, and we get

\begin{eqnarray}
G_{out} &\equiv& P_0\int r_cdr_cf(r_c) \int d\phi  e^{(V+i q r_c) \cos \phi} \int d\e\cosh \e 
e^{-(U-i Q_0 \tau_f)\cosh \e} \nonumber
\\
&=& 4\pi P_0 \int r_cdr_cf(r_c) I_0(V+iqr_c) K_1(U-iQ_0\t ).
\end{eqnarray}

\section{The case where $P_0=(p_{10}+p_{20})/2$}

In this case the formulae for $R^2_H$ are different from those given in Section~5 because the variable $U'\equiv \beta P_0  \cosh\th_c $ depends on $Q$. Indeed
\ba
P_0\equiv \frac12(p_{01}+p_{02})=\frac12\left[\sqrt{m^2+(\vec{P}
+\vec{Q}/2)^2}
+\sqrt{m^2+(\vec{P}-\vec{Q}/2)^2}\right].
\ea
Up to the second order in $Q$ we obtain 
\ba
P_0=\sqrt{m^2+P^2}\left[1+\frac{Q^2}{8(m^2+P^2)} -\frac{(\vec{P}\cdot \vec{Q})^2}{8(m^2+P^2)^2}\right].
\ea
Consequently, up to the second order in $Q$ we have 
\ba
e^{-U'\cosh\e_c}=e^{-U\cosh\e_c}\left[1-q^2(U/2) \xi\cosh\e_c\right],
\ea
where $U=\beta \sqrt{m^2+P^2} \cosh\th_c$ is given by (\ref{UandV})  and
\ba
\xi_{long}=\frac1{4m^2};\;\;\;\xi_{side}=\frac1{4m_\perp^2};\;\;\;\xi_{out}=\frac{m^2}{4m_\perp^4}.
\ea
Repeating the  argument given in Appendix B we thus obtain
\ba
R_H^2=\left<\chi^2\right>-\left<\chi\right>^2+\xi \left< U\cosh\e_c\right> -\xi
\ea 
with
\ba
\left<U\cosh\e_c\right>= \beta m_\p\frac{\int r_c dr_c f(r_c) \cosh\th_c I_0(V) K_0''(U) }{\int r_c dr_c f(r_c) I_0(V) K_1(U) }
\ea  
and where the last term represents the contribution from the
 factor $P_0$ in front of  (\ref{FT2}).  Note that in this case the contribution from the denominator is always as calculated in the present paper and never put equal zero \cite{Lisa:2005dd}.

\end{document}